\documentstyle[preprint,aps]{revtex}
\begin{document}
\baselineskip=.55cm

\title{  Reply to the paper of V.D. Efros: \\
       Comments in some recent papers on the Hyperspherical
       approach in few--body systems}

\author{ A.~Kievsky$^1$, L.E. Marcucci$^2$, S.~Rosati$^{1,3}$, M.Viviani$^1$}
\address{ $^1$Istituto Nazionale di Fisica Nucleare, Piazza Torricelli 2,
          56100 Pisa, Italy }
\address{ $^2$Phys. Dept. Old Dominion University, Norfolk, VA 23529, USA}
\address{ $^3$Dipartimento di Fisica, Universita' di Pisa, Piazza Torricelli 2,
          56100 Pisa, Italy }

\maketitle

\bigskip

 About a year ago V.D. Efros submitted a paper containing criticisms
of our recent research activity on bound and scattering states
of $A=3$, $4$ nucleons to a well-known international scientific review.
That paper was not published
and this unpleasant discussion was then avoided. Unfortunately,
V.D. Efros still ``feels himself obliged to discuss this issue in public''
\cite{efrosII}. As a consequence, we are forced to reexamine the
essence of the controversy and of the comments contained in
ref. \cite{efrosII}. These are the motivations of the present
paper. 

Studies and calculations of few nucleon systems have been performed
since  long time using different techniques.
In particular, the  Hyperspherical Harmonic (HH) approach and Faddeev
theory have been widely utilized.
The theoretical grounds for both methods are now well
established and most recent work has been  concerned  mainly with
improving the accuracy of the numerical results.

A few years ago we started studying few nucleon systems using
correlated Hyperspherical Harmonic bases, i.e.\ the Jastrow correlated basis
CHH or pair correlated basis PHH functions. Only rather
recently~\cite{KVR93} results whose accuracy is close to that of the best
available techniques~\cite{method} have been obtained by
applying such correlated expansions (CHH or PHH) to the case of the
$A=3$ system interacting with realistic NN potentials.
In ref.~\cite{KVR93} the three--nucleon wave function is written as a sum of
Faddeev-like amplitudes, each amplitude is decomposed in channels labelled
by angular--spin--isospin quantum numbers and expanded in terms of
correlated HH functions.
This type of decomposition is quite standard, since it
has been applied by many authors in a variety of papers.
Consequently in ref.~\cite{KVR93} the authors did not claim
any original contribution to the procedure. On the other
hand, since the problem is to obtain a truly accurate numerical solution
of the problem, particular care must be taken in selecting the HH functions
to be included in order
 to reduce the numerical calculations
and obtain accurate results .

In our approach the expansion of the
wave function is performed in two steps.
 In each channel all the correlated HH functions are considered
corresponding to increasing values of the grand angular quantum number
$K$ until the  desired accuracy is reached. 
The number of channels is then increased until
complete stability is achieved for the calculated quantities. In this
procedure, we select the maximum grand angular
quantum number $K_{max}$  to have
quite different values for the different channels so as to reflect
its importance in the description of the wave function.  
The choice of the set of correlated HH functions is therefore 
rather straightforward, however the
important point is in  the adoption of suitable correlation factors. To
give an idea of this point, in the case of realistic NN potentials the
triton binding energy can be obtained with three accurate digits by
using only $12$ channels and $6$ correlated HH functions per channel. 
This differs
from the uncorrelated HH expansion as used in
refs.~\cite{efros73,demin73,mukh89,fomin81}
where all the basis elements of the
selected channels having $K \le K_{max}$ are included. It has to be
stressed that with the uncorrelated expansion quite large values of $K_{max}$ 
must be considered, as can be inferred from ref. \cite{KMRV97}.

This procedure was also followed when calculations
were performed using the (uncorrelated) HH expansion~\cite{KMRV97}.
Here again no claim to originality was advanced. The paper was a
technical one, the purpose being to show that it is also possible with the
uncorrelated HH basis to calculate the observables of the
$A=3$ bound state systems with great accuracy, superior to
the present experimental values available. It is important to notice that
the situation is quite different when $A>3$, where the selection of
the important channels and the HH components to be included in the
expansion becomes a difficult problem \cite{Demin77,VRK95,VRK98}.

The paper by  Demin et al.~\cite{efros73} was cited in
ref.~\cite{KMRV97} without
any particular reference to their selection of the HH
functions. Moreover, the calculations provided in
ref.~\cite{demin73,mukh89,fomin81}  were
not accurate enough to fully justify the proposed selection.
In fact, using  the  uncorrelated HH expansion, one has
to overcome the  difficulty of how to manage the very large number of
basis  functions required in the case of {\sl hard} interactions,
such as the MT(I-III) potential which has a $1/r$ repulsion,
or of realistic interactions such as the Argonne type potentials.
In ref.~\cite{KMRV97} very precise solutions were obtained for
potentials of this kind, with an accuracy of six  digits for the binding
energy  and four digits for other quantities of interest.
In that context, we did not deem it appropriate to cite
for comparison the  papers~\cite{demin73,mukh89,fomin81}
mentioned in the comments. The reason is that,
in the latter papers, the most advanced application of the HH basis
is contained in  the paper by Mukhtarova~\cite{mukh89}, where
a procedure similar to that later used by us in ref.~\cite{KMRV97}
was applied. However,  it is dangerous to refer to the
paper~\cite{mukh89} since something is wrong there.
Presumably, that is why ref.~\cite{mukh89}
is not cited in any of the papers presenting accurate solutions of
the three-nucleon problem (see, for example, ref.~\cite{method2} and
references cited therein). To be more explicit,
the calculated triton binding energy for the SSC(C) interaction
in ref.~\cite{mukh89} is quoted to be $B=7.608$ MeV, obtained by including
HH functions up to $K_{max}=34$ and with an extrapolated value of $7.65$
MeV. The mixed symmetry percentage is presented as $P_{s'}=1.15$\%.
Our corresponding estimates (not presented in \cite{KMRV97}
since the  SSC(C)  interaction is rather soft and does
not offer serious convergence problems) are $B=7.5385$ MeV and
$P_{s'}= 1.238$\%, respectively. The latter values confirm those already
obtained
in ref.~\cite{method} using the Faddeev technique.
It is evident that the calculation of  ref.~\cite{mukh89} is not
correct.

It may be worth noting that, as also stated in ref.~\cite{KMRV97},
there is no problem in our approach with the uncorrelated HH basis 
in including three-nucleon interaction terms  and in still obtaining
very accurate results. This would not be  possible using a numerical
technique such as that proposed in
refs.~\cite{efros73,demin73,mukh89,fomin81}. 

With regard to the treatment in refs.~\cite{KVR93,KVR95} of few-body reactions
in the framework of the HH approach,  it should be observed that our
technique  is based on correlated PHH functions. This is not a ``minor
difference'', as stated in ref. \cite{efrosII}, but it is a crucial
point since it is thereby possible to perform
$n-d$ and $p-d$ scattering calculations with great accuracy. As a matter of
fact, the
PHH and Faddeev  results have been successfully compared~\cite{huber95}
and proposed as a benchmark for different approaches to the problem.
On the other hand, the application  of the HH expansion
to scattering and reaction problems
with a view to obtaining accurate results so as
to make a fruitful comparison with the
corresponding experimental data is as yet  problematic.
In fact, the convergence rate of the uncorrelated HH expansion
for calculating $N-d$ scattering observables is for realistic
potentials even slower than that observed in the $A=3$ bound state
problem.

Moreover, the decomposition of the wave function as a sum of an
asymptotic and an internal part is ``natural'', as also stated by
V.D. Efros and coll. in ref. \cite{ZPE69}. In fact,  the authors
of  ref. \cite{ZPE69} did not deem it necessary to insert any specific
references to  earlier papers where
the wave function was  written in just that way. To be explicit,
that decomposition was already used in various papers
in the fifties and early sixties for studying $n-d$ scattering
(see for example ref. \cite{Hum65}).
The fact is that for a numerical application of the Kohn--Hulth\` en
variational principle, that decomposition of the trial wave function
is rather obvious. For example, a very detailed review of the status
at that time of the variational approach to $n-d$ scattering  is
reported in  ref. \cite{Delves}. There too, even though  posterior
to ref. \cite{ZPE69}, the  decomposition of the trial wave function is
reported without any  particular comment.  

To be noticed, that in our {\sl first} paper on $N-d$
scattering~\cite{KVR93}, we
have stated: ``The  variational approach based on the use of PHH
correlated functions can be  extended to investigate scattering states
and in this section the  application to the N--d scattering below the
break--up threshold is discussed.  Following the pioneering work of
Delves~\cite{Delves} for realistic NN  interactions, the wave function
for a N--d scattering state will be written as
\begin{equation}
   \Psi = \Psi_{C} + \Psi_{A}\ .\label{eq:scatte}
\end{equation}
The first term $\Psi_{C}\ldots$'', etc. Therefore, it is 
``misleading'' that the author of ref. \cite{efrosII} reported in
his comment a citation of our subsequent paper~\cite{KVR95} 
(which, moreover,  is  a Rapid Communication and therefore 
a rather short paper).

Last but not least the following remark appears to be worthy of
consideration. A number of papers by the present authors devoted to
the study of bound and scattering states of three and four nucleons have been
published in international reviews. None of the referees has ever
asked us to add any of the references cited by V.D. Efros.


\bigskip


\begin{thebibliography}{99}
\bibitem{efrosII} V.D. Efros, nucl-th/9804036.
\bibitem{KVR93}A. Kievsky, M. Viviani and S. Rosati, Nucl. Phys.
  {\bf A 551} (1993) 241.
\bibitem{method}C.R. Chen, G.L. Payne, J.L. Friar and B.F. Gibson, Phys. Rev.
  {\bf C31} (1985) 266; T. Sasakawa and S. Ishikawa, Few--Body Syst.
  {\bf 1} (1986) 3;
  H. Kameyama, M. Kamimura and Y. Fukushima, Phys. Rev. {\bf C40}
  (1989) 974; W. Gl\" ockle {\sl et al.}, Few--Body Systems Suppl. {\bf
  8} (1995) 9.
\bibitem{efros73} V.F. Demin, Yu. E. Pokrovsky and V.D.  Efros,
        Phys. Lett. {\bf B44} (1973) 227.
\bibitem{demin73} V.F. Demin and Yu.E.  Pokrovsky,
        Phys. Lett. {\bf B47} (1973) 394.
\bibitem{mukh89} M.I. Mukhtarova, Sov. J. Nucl. Phys. {\bf 49} (1989) 208.
\bibitem{fomin81} B.A. Fomin and V.D.  Efros, Sov. J. Nucl. Phys. {\bf 34}
 (1981) 587; Phys. Lett. {\bf B98} (1981) 389.
\bibitem{KMRV97} A. Kievsky, L.E. Marcucci, S. Rosati and M. Viviani,
        Few-Body Systems {\bf 22} (1997) 1.
\bibitem{Demin77} V.F. Demin, Sov. J. Nucl. Phys. {\bf 26} (1978) 379.
\bibitem{VRK95} M. Viviani, A. Kievsky and S. Rosati, Few--Body
                Systems {\bf 18} (1995) 25.
\bibitem{VRK98} M. Viviani, S. Rosati and A. Kievsky, in
                preparation.
\bibitem{method2} K. Varga and Y. Suzuki, Phys. Rev. {\bf 52}
                 (1995) 2885; K. Varga, Y. Ohbayasi and Y. Suzuki,
                  Phys. Lett. {\bf B396} (1997) 1.
\bibitem{KVR95}A. Kievsky, M. Viviani and S. Rosati, Phys. Rev. {\bf C52}
               (1995) R15.
\bibitem{huber95}D. Hueber {\sl et al.}, Phys. Rev. {\bf C51} (1995) 1100.
\bibitem{ZPE69} B.N. Zakhar'ev, V.V. Pustolatov and V.D. Efros,
Sov. J. Nucl. Phys. {\bf 8} (1969) 234.
\bibitem{Hum65} J. W. Humberston, Nucl. Phys. {\bf 69} (1965) 291.
\bibitem{Delves}L.M. Delves, Advances in Nuclear Physics, Vol.5,
          ed. M. Baranger and E. Vogt (New York-London: Plenum Press 1972),
          p.126
\end{thebibliography}
\end{document}